\documentclass[12pt]{iopart}

\usepackage{graphicx}

\begin{document}

\title{Yang-Baxter integrable models in experiments: from condensed matter to ultracold atoms}

\author{Murray T. Batchelor$^{1,2}$ and Angela Foerster$^{3}$}

\address{$^{1}$ Centre for Modern Physics, Chongqing University, Chongqing 400044, China}
\address{$^{2}$ Department of Theoretical Physics,
Research School of Physics and Engineering, and Mathematical Sciences Institute, 
Australian National University, Canberra ACT 0200, Australia}
\address{$^{3}$ Instituto de F\'{\i}sica da UFRGS, 
Av. Bento Gon\c{c}alves, 9500, Agronomia, Porto Alegre - RS - Brazil}

\ead{batchelor@cqu.edu.cn}
\ead{angela@if.ufrgs.br}

%\date{}
%\submitto{\jpa}
%\maketitle

\begin{abstract}
The Yang-Baxter equation has long been recognised as the masterkey to integrability, providing the basis for 
exactly solved models which capture the fundamental physics of a number of 
realistic classical and quantum systems. 
In this article we provide an introductory overview of the impact of Yang-Baxter integrable models on experiments 
in condensed matter physics and ultracold atoms. 
A number of prominent examples are mentioned, including the hard-hexagon model, the Heisenberg spin chain, the transverse 
quantum Ising chain, a spin ladder model, the Lieb-Liniger Bose gas, the Gaudin-Yang Fermi gas and the two-site 
Bose-Hubbard model.
The review concludes by pointing to some other recent developments with promise for further progress. 
\end{abstract}

\tableofcontents

\section{Introduction}

The purpose of this review article is to provide an introductory overview of the impact of exactly solved models, 
more specifically, the impact of Yang-Baxter integrable models, on experiments. 
The usefulness of areas of study like statistical mechanics and quantum field theory, 
which were ultimately seen to be deeply related, is that they address fundamental problems 
involving realistic model systems of interacting states or particles.
In addition to elegant exact solutions of key models, the beauty of these subjects is that mathematical methods 
developed to tackle the problems at hand have flowered in their own right.
A striking example is the Yang-Baxter equation, which has inspired and led to remarkable developments in 
different branches of mathematics. 
For example, Yang-Baxter integrable models provide explicit realisations of algebraic structures, such as Lie algebras and quantum groups.
On the physics side, both Onsager's exact solution of the two-dimensional Ising model and Baxter's 
exact solution of the eight-vertex model played key roles in the development of the modern theory of 
phase transitions and critical phenomena. 
There are several forces at play in these developments. 
As already alluded to, the models themselves are fundamental and nontrivial. 
It should thus come as no surprise that they might eventually be realised in hitherto unimagined experimental settings. 
However, this was not of course the original motivation for much of this work, which has spanned many decades.  
Rather, the {\em raison d' \^{e}tre} was explained by Baxter himself in 1982 \cite{Baxter}:

\begin{quote}
{\it Basically, I suppose the justification for studying these 
models is very simple:  they are relevant and they can be solved, so why not do so and see what 
 they tell us?}
\end{quote}

And the models have told us so much! 
Exactly solved models of this kind can be found in many areas of physics, such as
condensed matter, quantum field theory, the AdS/CFT correspondence in string theory, 
nuclear physics, atomic and molecular physics and ultracold atoms.
Indeed, the Yang-Baxter equation was quite early recognised as the masterkey to integrability.\footnote{For 
an historical perspective, including the different forms of the Yang-Baxter equation, see, e.g., \cite{Perks}.}
Obtaining a solution to the Yang-Baxter equation is tantamount to finding an exactly solved model.
Here we present some prominent, and rather selective, examples of Yang-Baxter integrable models 
which have been realised in experiments 
in the areas of condensed matter physics and ultracold atoms.\footnote{Some of these developments have been 
highlighted elsewhere recently \cite{Batchelor,Guan}.}
Other related developments and some future prospects are mentioned in the conclusion.

\section{Yang-Baxter integrable models in condensed matter}

Condensed matter physics has long been the traditional setting in which a range of  
exactly solved models have been realised in experiments. 
These include the early experiments related to the two-dimensional Ising model.
In this section we discuss the hard hexagon model, 
the Heisenberg spin chain, 
the quantum Ising chain with transverse and longitudinal fields, 
and a spin ladder model.

\subsection{Hard hexagon model}

The hard-hexagon model is a model defined on the triangular lattice, 
upon which hexagonal tiles are placed such that the tiles do not overlap.
Defining $g(n,N)$ as the number of ways of placing $n$ hexagons on $N$ sites, the 
combinatorial problem is to calculate 
\begin{equation}
f(z) = \lim_{n\to\infty} \left( \frac{1}{N} \ln Z_N \right), 
\end{equation}
with the grand partition function $Z_N$ defined by 
\begin{equation}
Z_N = \sum_{n=0}^{N/3} z^n g(n,N) .
\end{equation}
The variable $z$ is the fugacity, related to the density of hexagon tiles by $\rho = \partial f / \partial z$.
There is a transition from a dense solid phase to a low density ``fluid'' phase at $z=z_c$.
Baxter \cite{hh} obtained the result 
\begin{equation}
f(z) \sim  | z - z_c |^{2-\alpha},
\end{equation}
with critical exponent $\alpha=\frac13$ and critical point $z_c = \tau^5$, where $\tau=(1+\sqrt5)/2$.
The remarkable exact solution of this problem featured the Yang-Baxter equation and saw the first appearance of 
the celebrated Rogers-Ramanujan identities in statistical mechanics.\footnote{For a personal account of these developments 
the reader is referred to \cite{Baxter2015}. See also \cite{Barber}.}

%%%%%%%%%%%%%%%%%%%%%%%%%%%%%%%%%%%%%%%%%%%%%%%%%%%%%%%%%%%%%%%%%%%%%%%
\begin{figure}[t]
\begin{center}
\includegraphics[width=0.5\columnwidth]{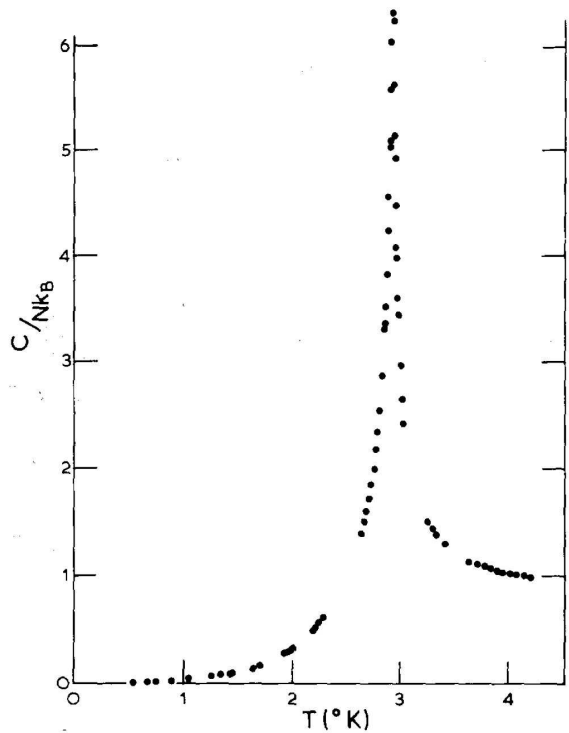}
\caption{Specific heat of helium on graphite at $\frac13$ coverage \cite{Bretz0,Bretz}. 
Version from \cite{Barber}.
}
\label{helium}
\end{center}
\end{figure}
%%%%%%%%%%%%%%%%%%%%%%%%%%%%%%%%%%%%%%%%%%%%%%%%%%%%%%%%%%%%%%%%%%%%%%%

There have been numerous experimental studies of the behaviour of thin films of gases adsorbed on regular crystal surfaces \cite{review}. 
In particular, the specific heat of single monolayers of helium adsorbed on graphite was measured \cite{Bretz0,Bretz}, with the   
spectacular peak characteristic of a second order phase transition shown in figure \ref{helium}. 
In the vicinity of the peak the specific heat is estimated \cite{Bretz} to diverge as
\begin{equation}
C \sim  | T - T_c |^{-\alpha},
\end{equation}
with exponent $\alpha \approx 0.36$. 
The transition observed is from a disordered arrangement of helium atoms at high temperature 
to an ordered state in which the helium atoms form a regular triangular arrangement.
Assuming that the helium atoms can only sit at the centres of the hexagons (which is an approximation) 
results in a model of a lattice gas on the triangular lattice.
The important observation is that the original lattice consists of three sublattices, 
only one of which can be completely occupied by helium atoms.
It was concluded that the transition is in the same universality class as the transition 
in the 3-state Potts model \cite{Alexander}.
In particular, from the equivalence $z \sim T$ Baxter's exact results for hard hexagons imply also that $\alpha=\frac13$ 
for the 3-state Potts model. 
One can thus see that the experimental estimate is rather accurate.

\subsection{Heisenberg spin chain}

The spin-$\frac12$ Heisenberg chain \cite{Heisenberg}
\begin{equation}
{\cal{H}}  = J  \sum_{j} {\vec{S}}_j \cdot \vec{S}_{j+1}, 
\end{equation}
defined in terms of spin-$\frac12$ operators ${\vec{S}}_j=(S_j^x,S_j^y,S_j^z)$ acting at site $j$, 
is the canonical example of a one-dimensional quantum many-body system solved exactly by 
means of the Bethe Ansatz \cite{Bethe,Gaudin}. 
Only much later was it realised that the underlying integrability of the Heisenberg chain is due to the Yang-Baxter equation, 
governed by the $R$-matrix of the related six-vertex model \cite{Baxter,Perks,Gaudin}.

\begin{figure}[t]
\begin{center}
\vskip 3mm
\includegraphics[width=0.8\columnwidth]{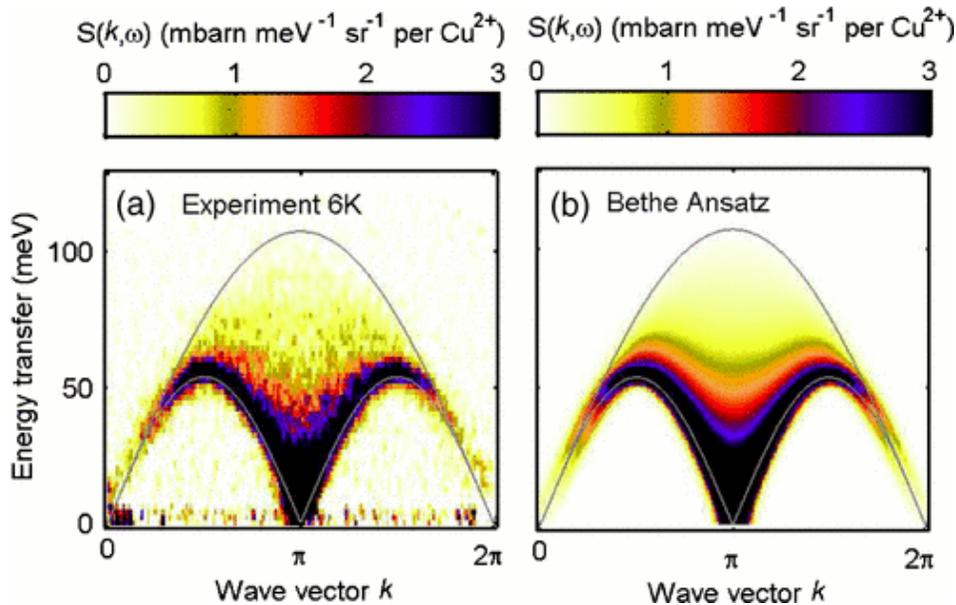}
\caption{Comparison between experiment and theory for the dynamical structure factor of the prototypical Heisenberg 
antiferromagnetic spin-$\frac12$ chain compound KCuF$_3$.
(a) Inelastic neutron scattering data from the ISIS Facility, Rutherford Appleton Laboratory, U.K. 
(b) Theoretical results computed via the algebraic Bethe Ansatz. 
From \cite{Lake}.
}
\label{XXZfig}
\end{center}
\end{figure}

In the compound KCuF$_3$ orbital order provides strong Heisenberg coupling between Cu$^{2+}$ spin-$\frac12$ ions.  
For energies above a certain threshold, the behaviour of KCuF$_3$ is entirely one-dimensional, 
making it an ideal testbed for the experimental study of the antiferromagnetic spin-$\frac12$ Heisenberg chain. 
The energy and wave vector dependence of the characteristic spinon continuum and 
the presence of universal scaling behaviour indicating proximity to the Luttinger liquid quantum critical point 
were established for the first time in KCuF$_3$. 
Figure~\ref{XXZfig} shows the direct comparison between the dynamical structure factor obtained from high quality 
inelastic neutron scattering data 
for the compound KCuF$_3$ and theoretical results obtained via the algebraic Bethe Ansatz for the 
antiferromagnetic spin-$\frac12$ Heisenberg chain \cite{Lake}.
This work also highlighted the inadequacy of conventional approximations in calculating the dynamical structure factor of the 
antiferromagnetic spin-$\frac12$ Heisenberg chain.

%%%%%%%%%%%%%%%%%%%%%%%%%%%%%%%%%%%%%%%%%%%%%%%%%%%%%%%%%%%%%%%%%%%%%%%
%%%%%%%%%%%%%%%%%%%%%%%%%%%%%%%%%%%%%%%%%%%%%%%%%%%%%%%%%%%%%%%%%%%%%%%

The lower and upper boundaries in the multispinon continuum 
in figure~\ref{XXZfig}(b) are defined by the curves  $\omega_l(k) = \frac{\pi}{2} J | \sin k|$ and 
$\omega_u(k) = {\pi}J | \sin \frac{k}{2}|$, with $\omega_l(k)$ describing the dispersion of the basic 
Faddeev-Takhtajan spinons carrying fractional spin-$\frac12$ \cite{FT}.

\subsection{Quantum Ising chain with transverse and longitudinal fields}

In a state-of-the-art experiment, a quasi-one-dimensional Ising ferromagnet was realised in CoNb$_2$O$_6$ (cobalt niobate) and 
tuned through its quantum critical point using strong transverse magnetic fields \cite{Coldea}. 
This is the quantum Ising chain with transverse and longitudinal fields -- the one-dimensional quantum counterpart 
of the two-dimensional classical Ising model in a magnetic field -- with hamiltonian 
\begin{equation}
H = - J \sum_j \left( S^z_j S^z_{j+1} + h \, S_j^x + h_z \, S_j^z \right) .
\end{equation}
This model has a quantum critical point at $h=h_c = J/2$ for $h_z=0$. 
In the scaling limit sufficiently close to the quantum critical point, i.e., $h_z \ll J, h=h_c$, 
the spectrum is expected to be described by Zamolodchikov's $E_8$ mass spectrum \cite{Zam}.
This is the remarkable Yang-Baxter integrable quantum field theory containing eight massive 
particles with a reflectionless factorized $S$-matrix.
Up to normalisation, the masses $m_i$ of these particles coincide with the 
components of the Perron-Frobenius vector of the 
Cartan matrix of the Lie algebra $E_8$.

\begin{figure}[t]
\begin{center}
\includegraphics [width=0.8\linewidth]{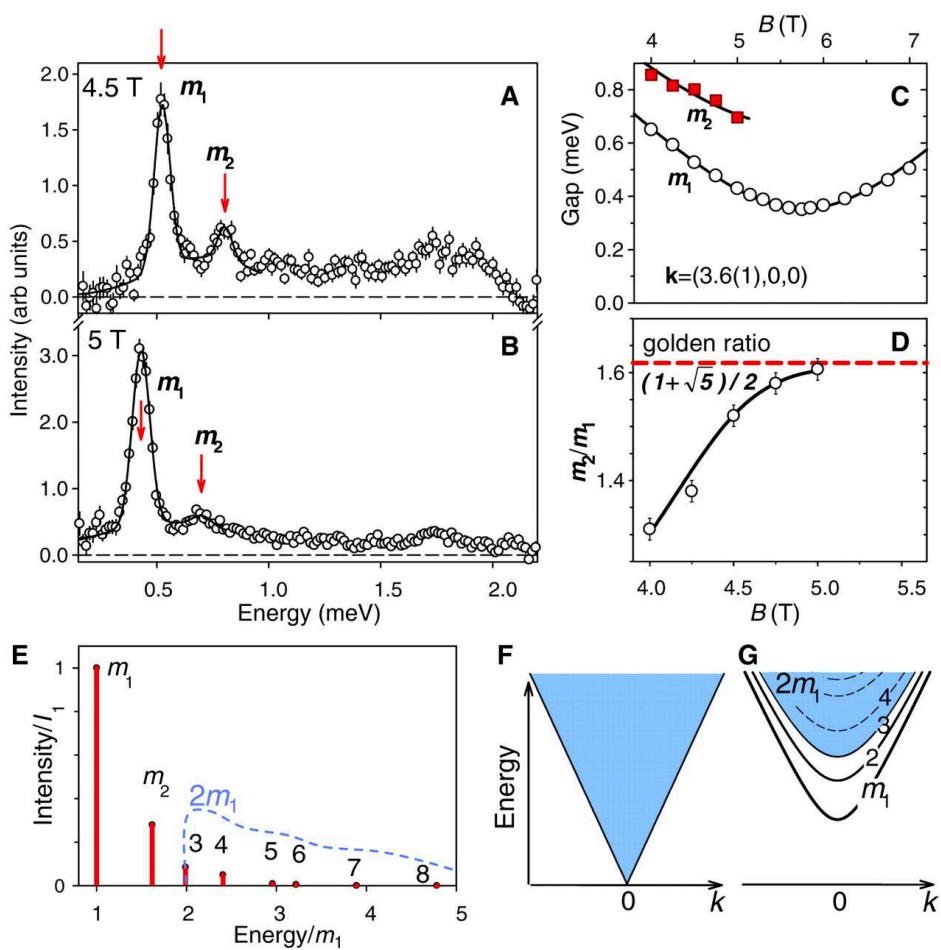}  
\end{center}
\caption{Various plots showing experimental evidence for the first few masses of 
the $E_8$ mass spectrum in the quasi-1D Ising ferromagnet CoNb$_2$O$_6$. 
From \cite{Coldea}. 
}
\label{massfig}
\end{figure}

The spectrum of this compound was observed by neutron scattering \cite{Coldea}, with clear 
evidence for the first few $E_8$ masses, see \fref{massfig}.
The emergence of such an exotic symmetry as $E_8$ has thus been observed in the lab.
We know that the integrable theory provides many more exact predictions \cite{aldo} than
experiments have been able to test so far.
Further developments are eagerly awaited on the experimental side.

%%%%%%%%%%%%%%%%%%%%%%%%%%%%%%%%%%%%%%%%%%%%%%%%%%%%%%%%%%%%%%%%%%%%%%%
%%%%%%%%%%%%%%%%%%%%%%%%%%%%%%%%%%%%%%%%%%%%%%%%%%%%%%%%%%%%%%%%%%%%%%%

\subsection{Spin ladder model}

\begin{figure}[ht]
\begin{center}
\includegraphics[width=0.45\linewidth]{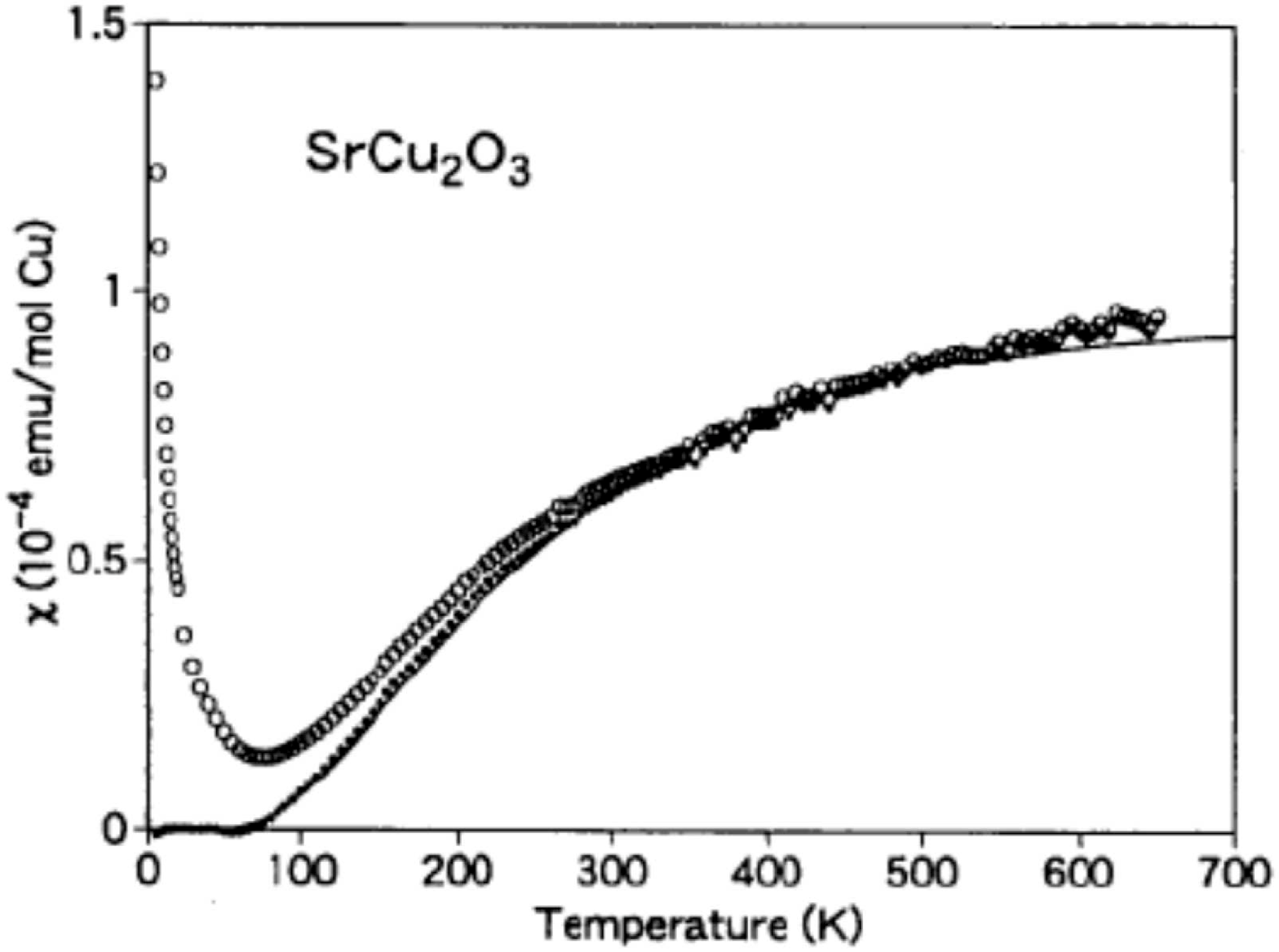}  
\includegraphics[width=0.44\linewidth]{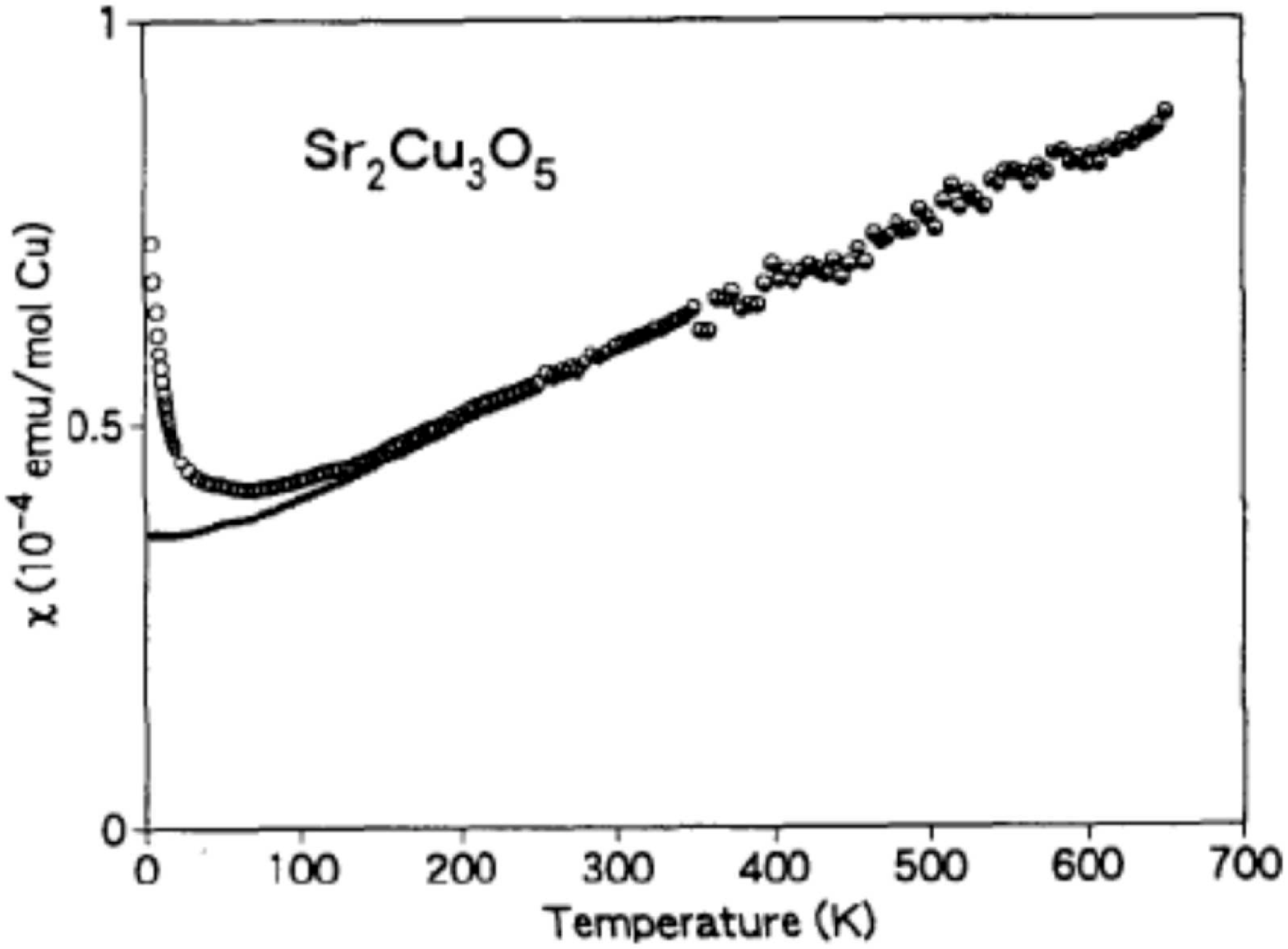}  
\end{center}
\caption{Susceptibility versus temperature for (left) the 2-leg ladder compound SrCu$_2$O$_3$ 
and (right) the 3-leg ladder compound Sr$_2$Cu$_3$O$_5$. From \cite{Azuma}.
}
\label{gaps}
\end{figure}

Many compounds have also been found with effective spin-$\frac12$ Heisenberg interactions between spins on a ladder-like structure. 
The physical properties of such spin ladders are dependent on the number of legs in the ladder. 
This different behaviour can be seen in figure~\ref{gaps} which shows the measured magnetic susceptibility curves as a 
function of temperature for the two-leg ladder compound SrCu$_2$O$_3$ and the three-leg ladder compound Sr$_2$Cu$_3$O$_5$.
In general a spin gap exists in the energy spectrum for even leg ladders while odd leg ladders do not exhibit a gap 
(see, e.g., \cite{Dagotto,ladder_review}).

%%%%%%%%%%%%%%%%%%%%%%%%%%%%%%%%%%%%%%%%%%%%%%%%%%%%%%%%%%%%%%%%%%%%%%%
%%%%%%%%%%%%%%%%%%%%%%%%%%%%%%%%%%%%%%%%%%%%%%%%%%%%%%%%%%%%%%%%%%%%%%%

An integrable spin ladder with two legs was introduced by Yupeng Wang \cite{wang} with Hamiltonian
\begin{equation}
{\cal H}=\frac{J_{\parallel}}{\gamma}\,{\cal H}_{{\rm leg}}+J_{\perp}\sum_{j=1}^{L}\vec{S}_j \cdot \vec{T}_j -
\mu_Bg H\sum_{j=1}^{L}(S_j^z + T^z_j), 
\end{equation}
where
\begin{equation}
{\cal H}_{{\rm leg}} = \sum_{j=1}^{L}\left(\vec{S}_j \cdot \vec{S}_{j+1}
+\vec{T}_j \cdot \vec{T}_{j+1}+ 4\, (\vec{S}_j \cdot \vec{S}_{j+1})(  \vec{T}_j  \cdot \vec{T}_{j+1}) \right) .
\end{equation}
Here $\vec{S}_j$ and $\vec{T}_j$ are spin-$\frac12$ operators acting on sites $j$ on the left and right legs of the ladder, 
$J_{\parallel}$ and $J_{\perp}$  are the leg and rung couplings, $\gamma$ is a 
rescaling constant, $H$ is the magnetic field and $L$ is the number of rungs.

\begin{figure}[ht]
\begin{center}
\includegraphics[width=0.49\linewidth]{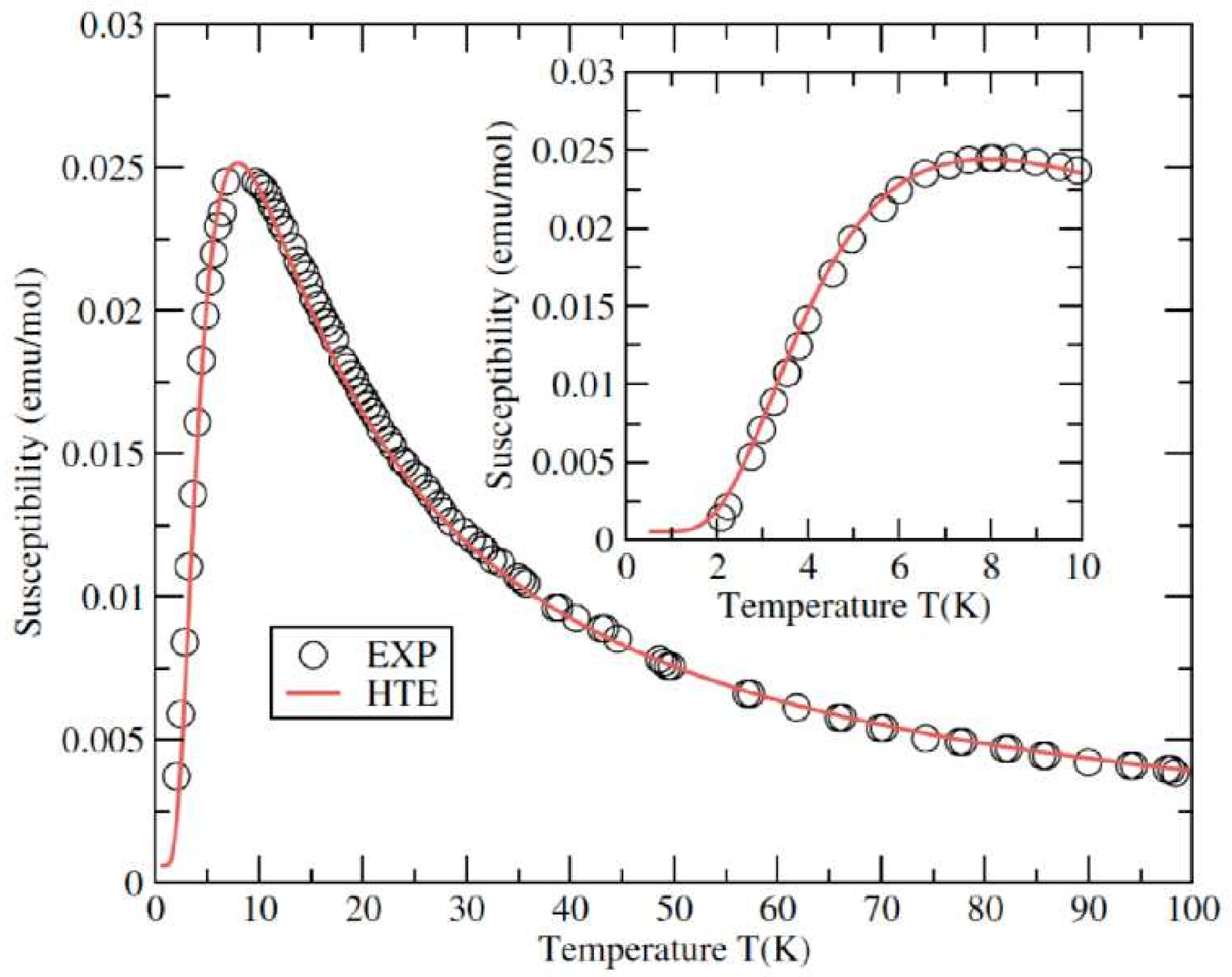}  
\includegraphics[width=0.49\linewidth]{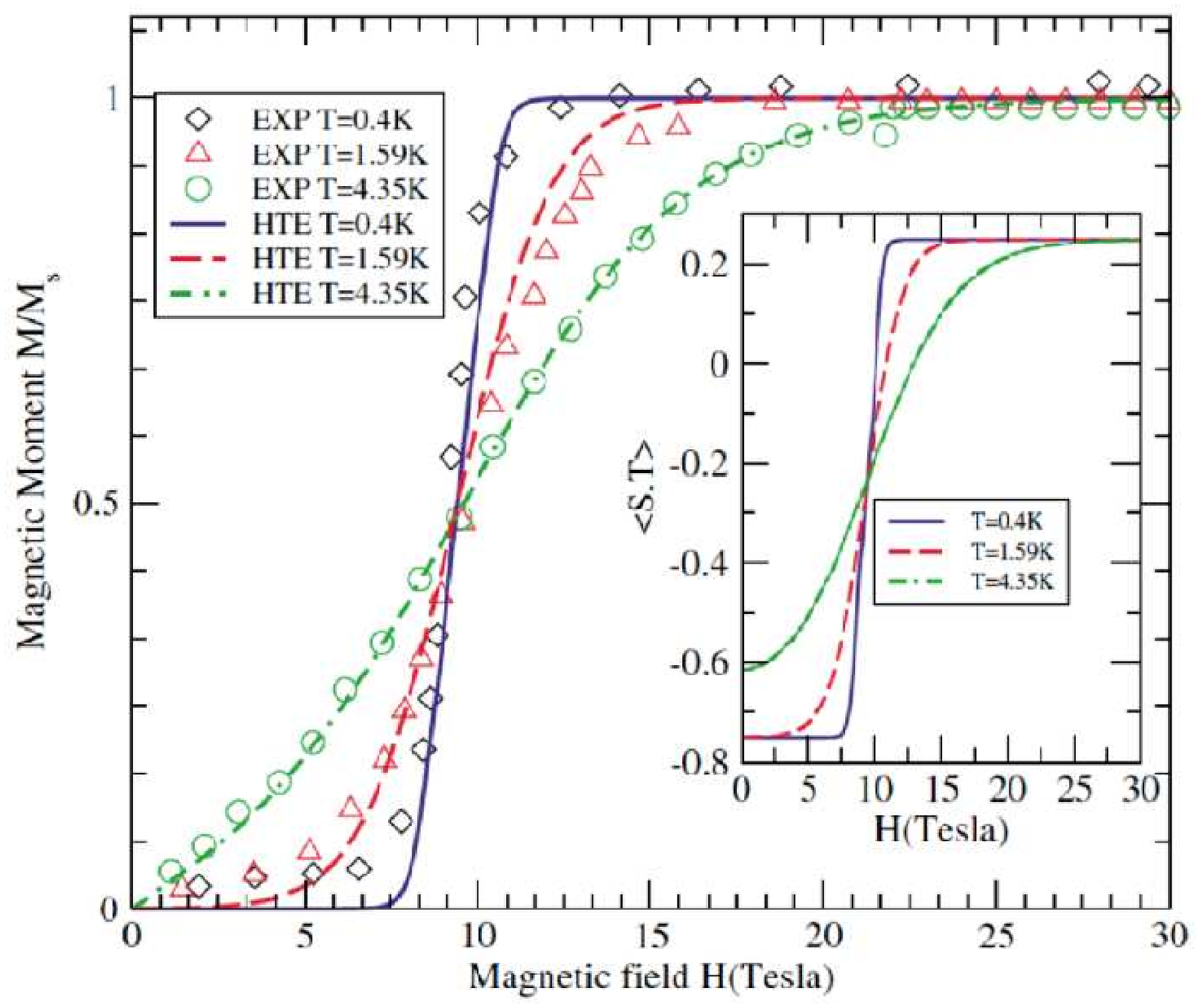}  
\end{center}
\caption{Comparison between experiment (EXP) and theory for the two-leg ladder compound (5IAP)$_2$CuBr$_4$$\cdot$2H$_2$O. 
(left) Susceptibility versus temperature at $H=1$T \cite{Landee}. 
The solid line denotes the susceptibility evaluated directly from the high temperature expansion (HTE) obtained from the 
exact solution. A parameter fit suggests the coupling constants $J_{\perp}=13.3$K and $J_{\parallel}=1.15$K with $\gamma =4$, 
$g=2.1$ and $\mu_B=0.672$K/T. The inset shows the same fit to the susceptibility at low temperature. 
(right) Magnetisation versus magnetic field \cite{Landee} with the same constants for different values of the temperature. 
The inset shows the one-point correlation function versus magnetic field. 
At $T=0.4$K the HTE magnetisation curve indicates the critical field values $H_{c1} \approx 8.3$T and $H_{c2} \approx 10.5$T, which are 
in excellent agreement with the experimental estimates $8.3$T and $10.4$T \cite{Landee}. 
From \cite{ladder_paper}.
}
\label{ladder1}
\end{figure}

This model differs from the usual Heisenberg spin ladder by the presence of the biquadratic interaction terms, which are 
in essence the price paid to ensure integrability.
The model is solved exactly in terms of the Bethe Ansatz: 
the leg part is simply the permutation operator corresponding to the $su(4$) algebra 
and the rung term becomes diagonal after a change of basis.
The gap and critical fields can be derived using the Thermodynamic Bethe Ansatz and the 
thermal and magnetic properties can be obtained using the Quantum Transfer Matrix method \cite{ladder_review,ladder_paper}.
This integrable ladder model can be used to accurately describe the physics of strong coupling ladder compounds, 
as illustrated in figure \ref{ladder1} for the two-leg ladder compound (5IAP)$_2$CuBr$_4$$\cdot$2H$_2$O.
Similar agreement is seen for other two-leg ladder compounds \cite{ladder_review,ladder_paper}.

%%%%%%%%%%%%%%%%%%%%%%%%%%%%%%%%%%%%%%%%%%%%%%%%%%%%%%%%%%%%%%%%%%%%%%%
%%%%%%%%%%%%%%%%%%%%%%%%%%%%%%%%%%%%%%%%%%%%%%%%%%%%%%%%%%%%%%%%%%%%%%%

\section{Yang-Baxter integrable models in ultracold atoms}

The experimental breakthroughs of trapping and cooling atoms to form Bose-Einstein \cite{break1} and fermionic \cite{break2} 
condensates led to 
extraordinary progress in the controlled study of a range of physical phenomena which had not been fully accessible, or at best 
only partially accessible, through experiments in the more traditional setting of condensed matter physics.
Further progress in confining the atoms in tight one-dimensional waveguides, with the ability to vary the interaction strength between 
atoms, opened up other exciting possibilities.
In addition to exploring various aspects of the physics of collective phenomena, the realisation of effectively one-dimensional quantum systems 
paved the way for contact with some well known exactly solved models. 
This in turn inspired further theoretical progress.
These exciting developments for one-dimensional systems have been reviewed recently \cite{bosons,GBL,Guan2}.

\subsection{Lieb-Liniger Bose gas}

The hamiltonian of $N$ interacting spinless bosons of mass $m$ with point interactions on a line of length $L$  is 
\begin{equation}
 {\cal H}= -\frac{\hbar^2}{2m} \sum_{i = 1}^{N}  \frac{\partial ^2}{\partial x_i^2}+ 2\, c \sum_{1\leq i<j\leq N}\delta (x_i-x_j), 
\label{Hamb}
\end{equation}
where $x_i$ are the boson co-ordinates and $c$ is the interaction strength. 
This is the model solved by Lieb and Liniger \cite{Lieb} by means of the Bethe Ansatz wavefunction. 
The underlying two-body $S$-matrix has a scalar form \cite{Gaudin,McGuire}.
The Lieb-Liniger model can be obtained in a scaling limit of the spin-$\frac12$ XXZ Heisenberg chain (see, e.g., \cite{GH87}).

%%%%%%%%%%%%%%%%%%%%%%%%%%%%%%%%%%%%%%%%%%%%%%%%%%%%%%%%%%%%%%%%%%%%%%%
\begin{figure}[ht]
\begin{center}
\includegraphics [width=0.60\linewidth]{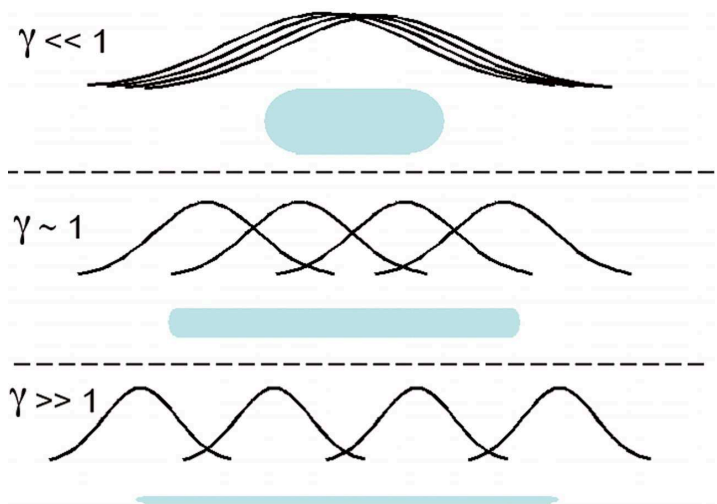}  
\end{center}
\caption{Cartoon showing the `fermionisation' of bosons as the interaction strength $\gamma$ is increased. 
For $\gamma \ll 1$ the behaviour is like a condensate, whereas for $\gamma \gg 1$ the behaviour is like the hard-core 
bosons of the Tonks-Girardeau gas. From \cite{KWW}.
}
\label{cartoonfig}
\end{figure}
%%%%%%%%%%%%%%%%%%%%%%%%%%%%%%%%%%%%%%%%%%%%%%%%%%%%%%%%%%%%%%%%%%%%%%%

\subsubsection{Repulsive regime.} 

In the analysis of the physics of this model it is convenient to define the dimensionless interaction parameter 
$\gamma = c/n$ in terms of the number density $n=N/L$.  
A cartoon of the expected atom distributions in the repulsive regime $c>0$, 
representing the `fermionisation' of the one-dimensional interacting Bose gas 
with increasing $\gamma$ is shown in \fref{cartoonfig}.

\begin{figure}[ht]
\begin{center}
\includegraphics [width=0.70\linewidth]{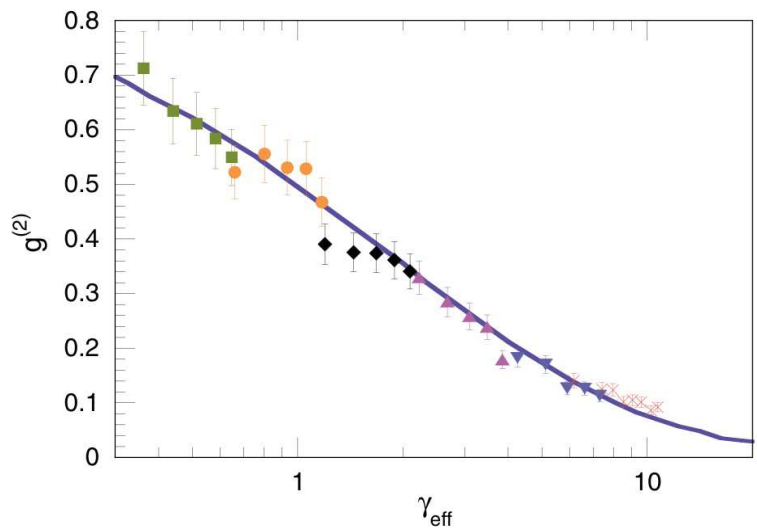}  
\end{center}
\caption{Local pair correlation function $g^{(2)}$ versus effective interaction strength obtained from measuring 
photoassociation rates in effective one-dimensional tubes of bosonic Rb atoms. 
The solid line is obtained from the Lieb-Liniger model. From \cite{KWWg2}.
}
\label{g2fig}
\end{figure}

One of the early experiments which made contact with the one-dimensional Lieb-Liniger model of interacting bosons 
measured local pair correlations in bosonic Rb atoms by photoassociation \cite{KWWg2}. 
The experimental measurement of the local pair correlation function $g^{(2)}$ is shown in \fref{g2fig}. 
This pair correlation function is proportional to the probability of observing two particles in the same location. 
As expected, the curve drops off towards zero as the interaction strength increases, just like for a 
non-interacting Fermi gas (recall \fref{cartoonfig}).

%%%%%%%%%%%%%%%%%%%%%%%%%%%%%%%%%%%%%%%%%%%%%%%%%%%%%%%%%%%%%%%%%%%%%%%
%%%%%%%%%%%%%%%%%%%%%%%%%%%%%%%%%%%%%%%%%%%%%%%%%%%%%%%%%%%%%%%%%%%%%%%

\subsubsection{Attractive regime.}

\begin{figure}[t]
\begin{center}
\includegraphics [width=0.7\linewidth]{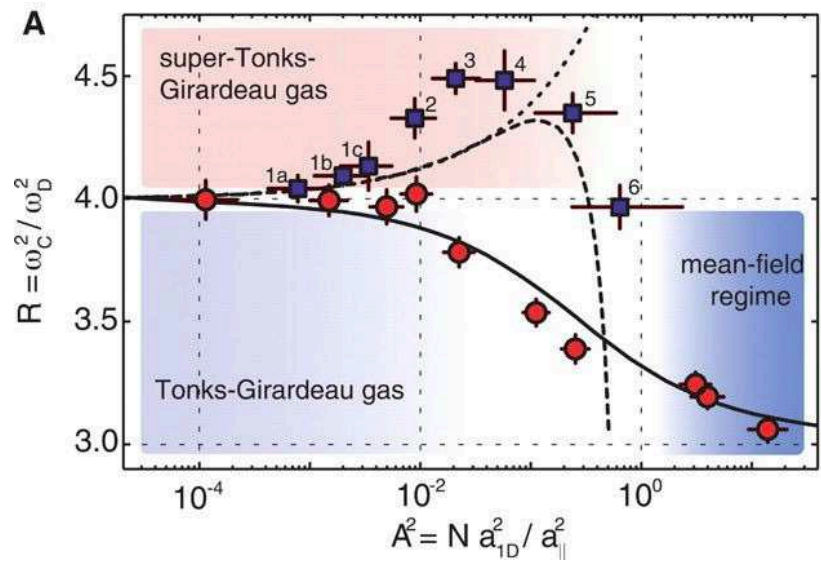}  
\end{center}
\caption{Plots of experimental data indicating the realisation of the one-dimensional super Tonks-Girardeau gas 
with bosonic cesium atoms. From \cite{Haller}. 
}
\label{sTGfig}
\end{figure}

The attractive regime $c<0$ has also been of interest. 
Inspired by Monte Carlo results which predicted the existence of a super Tonks-Girardeau gas-like state in the attractive interaction 
regime of quasi-one-dimensional Bose gases \cite{Astra}, it was shown that 
a super Tonks-Giradeau gas-like state corresponds to a highly-excited Bethe Ansatz state in the integrable Lieb-Liniger Bose gas 
with attractive interactions, for which the bosons acquire hard-core behaviour \cite{sTG}.
As the interaction is switched from strongly repulsive to strongly attractive 
the large kinetic energy inherited from the Tonks-Girardeau gas in a Fermi-pressure-like manner 
prevents the gas from collapsing.
Using a tunable quantum gas of bosonic cesium atoms, bosons in the attractive regime were realised and controlled in 
a one-dimensional geometry to obtain the super Tonks-Girardeau gas \cite{Haller}.
The highly excited quantum phase was stabilised in the 
presence of attractive interactions by maintaining and strengthening quantum correlations across a 
confinement-induced resonance (see \fref{sTGfig}).
This opened up the experimental study of metastable, excited, many-body phases with strong correlations.
In the super Tonks-Girardeau gas relaxation processes are suppressed, making the system metastable over long time scales \cite{PDC}.

%%%%%%%%%%%%%%%%%%%%%%%%%%%%%%%%%%%%%%%%%%%%%%%%%%%%%%%%%%%%%%%%%%%%%%%
%%%%%%%%%%%%%%%%%%%%%%%%%%%%%%%%%%%%%%%%%%%%%%%%%%%%%%%%%%%%%%%%%%%%%%%

\subsubsection{Excitations.}

Most recently the excitations of a strongly interacting one-dimensional bosonic quantum gas of $^{87}$Rb atoms have been probed 
experimentally at low temperature using Bragg spectroscopy and
compared to theoretical predictions based on the Lieb-Liniger model for the dynamical structure factor \cite{dyn}.
The excitation spectrum of an ultracold one-dimensional Bose gas of cesium atoms has also been probed experimentally 
using Bragg spectroscopy \cite{probe}.
The repulsive contact interactions are tuned from the weakly to the strongly interacting regime via a magnetic Feshbach resonance. 
The measured dynamical structure factor is compared to integrability-based calculations valid 
at arbitrary interactions and finite temperatures. 
The results highlight that hole-like excitations, which have no counterpart in higher dimensions and are a distinctive feature of the 
Bethe Ansatz analysis, actively shape the dynamical response of the gas \cite{probe}.

It is worth pointing out here in particular that the integrability of the Lieb-Liniger model is actually broken by the presence of 
the trapping potential along the axis of the one-dimensional gas. 
In order to compute quantitates like the dynamical structure factor of the inhomogeneous trapped gas, 
the way forward is to use the local density approximation (LDA), where
the response of the gas is assumed to be a sum of responses of small portions along the trap with different densities \cite{bosons,probe,LDA}. 
One then verifies, for example, that the response of the inhomogeneous gas is well approximated by the response of a uniform gas 
having a density equal to the mean density of the trapped gas. 
In general the integrable model still drives the underlying physics of the trapped gas.

\subsubsection{Interference experiments.}

Yang-Baxter integrability in cold gases has also been used to 
construct the full distribution function of contrasts for the interference pattern between two 
independent one-dimensional condensates \cite{Gritsev1}.
In particular, explicit use was made of Baxter's $Q$-operator \cite{Baxter}.
The interference-fringe contrast was successfully checked experimentally using two independent one-dimensional 
quantum degenerate atomic Bose gases of Rb$^{87}$ atoms 
created in a radio-frequency induced micro-trap on an atom chip \cite{Gritsev2}.
In addition to optical lattices, atom chips have provided another platform for 
realising one-dimensional Bose gases and various aspects of the Lieb-Liniger model \cite{bosons,Guan2}.

\subsection{Gaudin-Yang Fermi gas with polarization}

The Hamiltonian
\begin{equation}
{\cal H} = -\frac{\hbar^2}{2m}\sum_{i=1}^{N}\frac{\partial^2}{\partial x_{i}^2} + g_{1D}
\sum_{1\leq i<j\leq N} 
\delta(x_i - x_j) - \frac12 {H}(N-2M)
\end{equation}
describes $N$ spin-$\frac12$ fermions of mass $m$ interacting via a contact potential on a 
line of length $L$ and subject to an external magnetic field $H$. 
$M$ is the number of spin down fermions, with thus $N-M$ spin up fermions.
The interaction strength $g_{1D} ={\hbar^2c}/{m}$ can be tuned between the 
strongly attractive ($g_{1D}<0$) and strongly repulsive ($g_{1D}>0$) regimes. 
Here we consider the attractive regime.
This model was solved long ago by Gaudin \cite{Gaudin_f} and Yang \cite{Yang} in terms of the nested Bethe Ansatz,  
with a correspondingly more complicated $S$-matrix \cite{Yang,Smatrices,GBL}.
It received renewed interest in connection with ultracold atomic gases \cite{GBL,BFGK}.

%%%%%%%%%%%%%%%%%%%%%%%%%%%%%%%%%%%%%%%%%%%%%%%%%%%%%%%%%%%%%%%%%%%%%%%
\begin{figure}[ht]
\begin{center}
\includegraphics [width=0.8\linewidth]{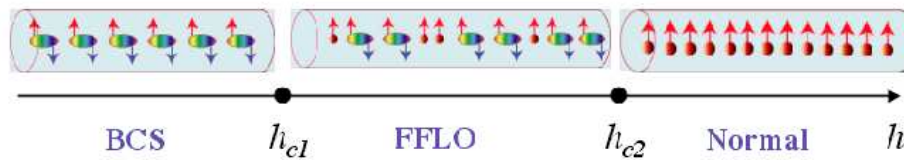}  
\end{center}
\caption{The zero temperature phase diagram of the Gaudin-Yang model 
as a function of magnetic field $h$ for given chemical potential. The three phases are the fully paired (BCS) phase, which is a quasi-condensate 
with zero polarization ($P = 0$), the fully polarized (Normal) phase with $P = 1$, and the partially polarized (FFLO) phase where $0 < P < 1$. 
The Fulde-Ferrell-Larkin-Ovchinnikov (FFLO) phase is a mixture of pairs and leftover (unpaired) fermions. 
The quantum critical points $h_{c1}$ and $h_{c2}$ separate the 
FFLO phase from the BCS phase and the normal phase. From \cite{ZL}.
}
\label{schematic}
\end{figure}
%%%%%%%%%%%%%%%%%%%%%%%%%%%%%%%%%%%%%%%%%%%%%%%%%%%%%%%%%%%%%%%%%%%%%%%

\begin{figure}[ht]
\begin{center}
\includegraphics [width=0.98\linewidth]{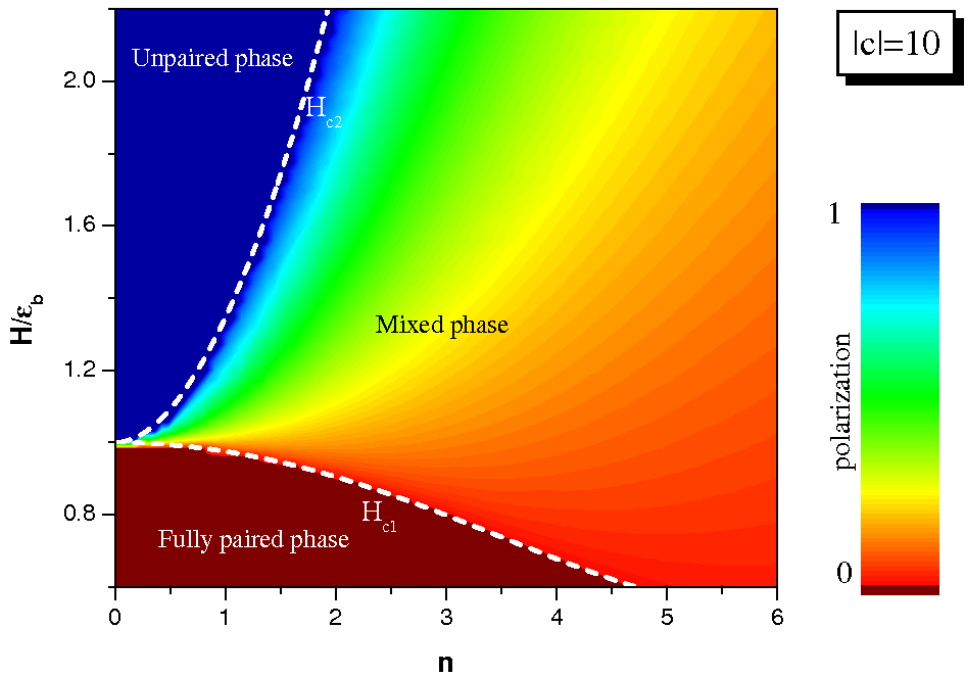}  
\end{center}
\caption{Phase diagram of the Gaudin-Yang model for strong coupling $|c| = 10$ in the $H-n$ plane, where 
$H$ is the external magnetic field and $n = N/L$ is the linear density.
The dashed lines are analytical results for the critical fields obtained from analysis of the dressed energy 
Thermodynamic Bethe Ansatz equations. These lines are to be compared with the 
phases (coloured by polarization) obtained from the numerical solution of the dressed energy equations.  
From \cite{HFGB}.
}
\label{colour}
\end{figure}

\begin{figure}[ht]
\begin{center}
\includegraphics [width=0.5\linewidth,angle=-90]{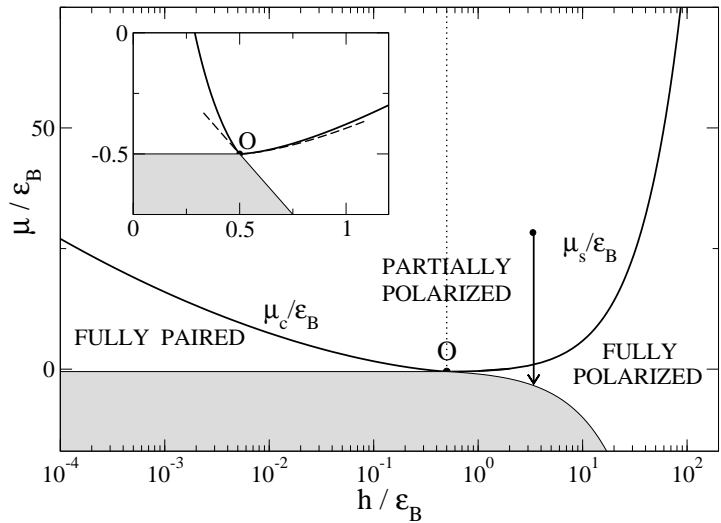}  
\end{center}
\caption{Phase diagram of the Gaudin-Yang model as a function of chemical potential 
and magnetic field. From \cite{Orso}.
}
\label{orsofig}
\end{figure}

Defining the polarization $P = (N-2M)/N$, the special case $M=N/2$ for 
which $P=0$ is known as the balanced case.
In the attractive regime the roots of the Bethe Ansatz equations tend to form pairs 
which can be broken by the magnetic field. 
The quantum critical points distinguishing the different quantum phases (see \fref{schematic}) 
can be calculated (see \fref{colour}) 
and the full phase diagram mapped out (see \fref{orsofig}).
It is clear that at $T=0$ the model predicts three phases in the strong coupling regime:
(i) a superfluid phase with zero polarization, where the ground state is composed of pairs; 
(ii) a fully polarized or ferromagnetic phase with full polarization and 
(iii) a partially polarized phase, where the ground state is a mixture of pairs and unpolarized fermions.

%%%%%%%%%%%%%%%%%%%%%%%%%%%%%%%%%%%%%%%%%%%%%%%%%%%%%%%%%%%%%%%%%%%%%%%
%%%%%%%%%%%%%%%%%%%%%%%%%%%%%%%%%%%%%%%%%%%%%%%%%%%%%%%%%%%%%%%%%%%%%%%

%%%%%%%%%%%%%%%%%%%%%%%%%%%%%%%%%%%%%%%%%%%%%%%%%%%%%%%%%%%%%%%%%%%%%%%
%%%%%%%%%%%%%%%%%%%%%%%%%%%%%%%%%%%%%%%%%%%%%%%%%%%%%%%%%%%%%%%%%%%%%%%

\begin{figure}[t]
\begin{center}
\includegraphics [width=0.8\linewidth]{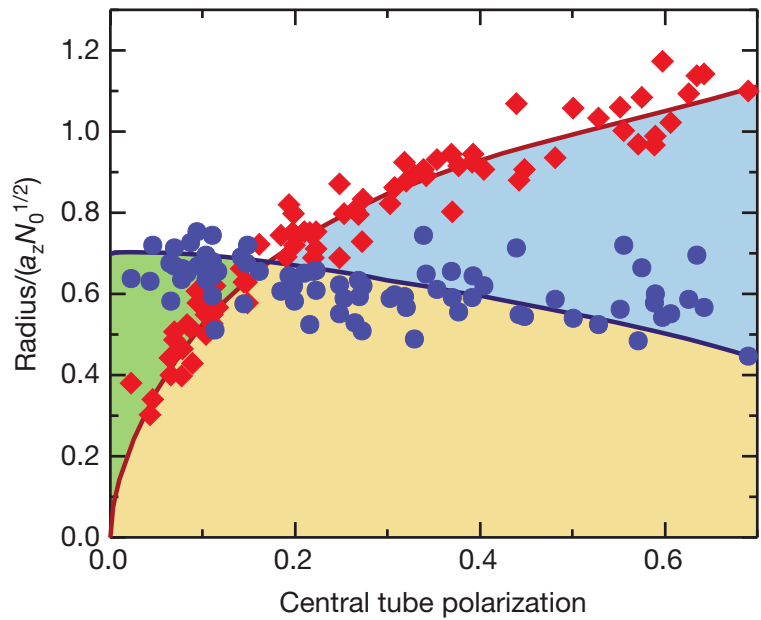}  
\end{center}
\caption{Experimental phase diagram of one-dimensional two-component fermions as a function of polarization. 
The red diamonds and blue circles denote the scaled radii of the axial density difference and the minority state axial density, 
respectively. The solid lines follow from the Gaudin-Yang model (recall figure \ref{orsofig}). From \cite{Rice}.
}
\label{fermionfig}
\end{figure}

The phase diagram was confirmed by experiments using fermionic $^6$Li atoms confined to one-dimension \cite{Rice}.
The system has attractive interactions with a spin population imbalance caused by a difference in the number 
of spin-up and spin-down atoms. 
Experimentally, the gas is dilute and strongly interacting.
The key features of the phase diagram (recall \fref{orsofig}) were experimentally confirmed using 
finite temperature density profiles (see \fref{fermionfig}).
The system has a partially polarized core surrounded by either fully paired or fully polarized wings at low temperatures, 
in agreement with theoretical predictions \cite{GBL,Rice}. 
More generally, the experimental results verified the coexistence of pairing and polarization at quantum criticality.

Just as various aspects of the one-dimensional Bose gas have been realised and studied in 
now over twenty different experiments \cite{bosons,Guan2}, 
it is anticipated that various aspects of one-dimensional fermions will be similarly tested and explored in the near future.
Recently one-dimensional quantum wires of repulsive fermions with a tunable number of spin components 
were realised by tightly trapping ultracold $^{173}$Yb atoms in a two-dimensional optical lattice \cite{SUN}.
Optical spin manipulation and detection techniques enabled the system to be prepared 
in an arbitrary number $N \le 6$ of spin components, thus realising different SU($N$) symmetries.
In the study of multi-component quantum systems Wilson ratios relating magnetic or particle fluctuations to 
thermal fluctuations 
have been shown to be particularly powerful dimensionless quantities for probing 
different magnetic phases in quantum liquids of this kind \cite{WR1,WR2}.

%%%%%%%%%%%%%%%%%%%%%%%%%%%%%%%%%%%%%%%%%%%%%%%%%%%%%%%%%%%%%%%%%%%%%%%
%%%%%%%%%%%%%%%%%%%%%%%%%%%%%%%%%%%%%%%%%%%%%%%%%%%%%%%%%%%%%%%%%%%%%%%

\subsection{Two-site Bose Hubbard model}

The two-site Bose-Hubbard model, also known as the canonical Josephson Hamiltonian \cite{leggett}, 
is a simple model used to study Josephson
tunnelling between two Bose-Einstein condensates, whose 
Hamiltonian is given by
\begin{equation}
H=\frac18 {K}(N_1-N_2)^2 - \frac12 {\Delta \mu}(N_1-N_2) - \frac12 {\cal E}_J (a^{\dag}_1 a_2 +a_2^{\dag} a_1), 
\end{equation}
Despite it's apparent simplicity the model captures the essence of competing linear and
nonlinear interactions, leading to interesting, nontrivial behaviour. 
It is not only relevant in the discussion of  tunnelling 
in  Bose-Einstein condensates, but also applicable to mesoscopic solid state 
Josephson junctions \cite{smerzi} and nonlinear optics \cite{optics}. 
Integrable cases of similar models with higher number of modes have been discussed in \cite{eric, arlei2}.
Concerning the notation, $a_1^\dagger, a_2^\dagger$ denote the single-particle creation
 operators in the two wells and  $N_1 = a_1^\dagger a_1,
 N_2 = a_2^\dagger a_2$ are the corresponding number operators. 
 The total boson number $N_1+N_2$
 is conserved and set to the fixed value of $N$.
 The coupling $K$ provides the strength of the scattering interaction between 
 particles,
 $\Delta \mu$ is the external potential and ${\cal E}_J$ is the coupling
 for the tunnelling.

 %%%%%%%%%%%%%%%%%%%%%%%%%%%%%%%%%%%%%%%%%%%%%%%%%%%%%%%%%%%%%%%%%%%%%%%
\begin{figure}[t]
\centering
\includegraphics [width=1.0\linewidth]{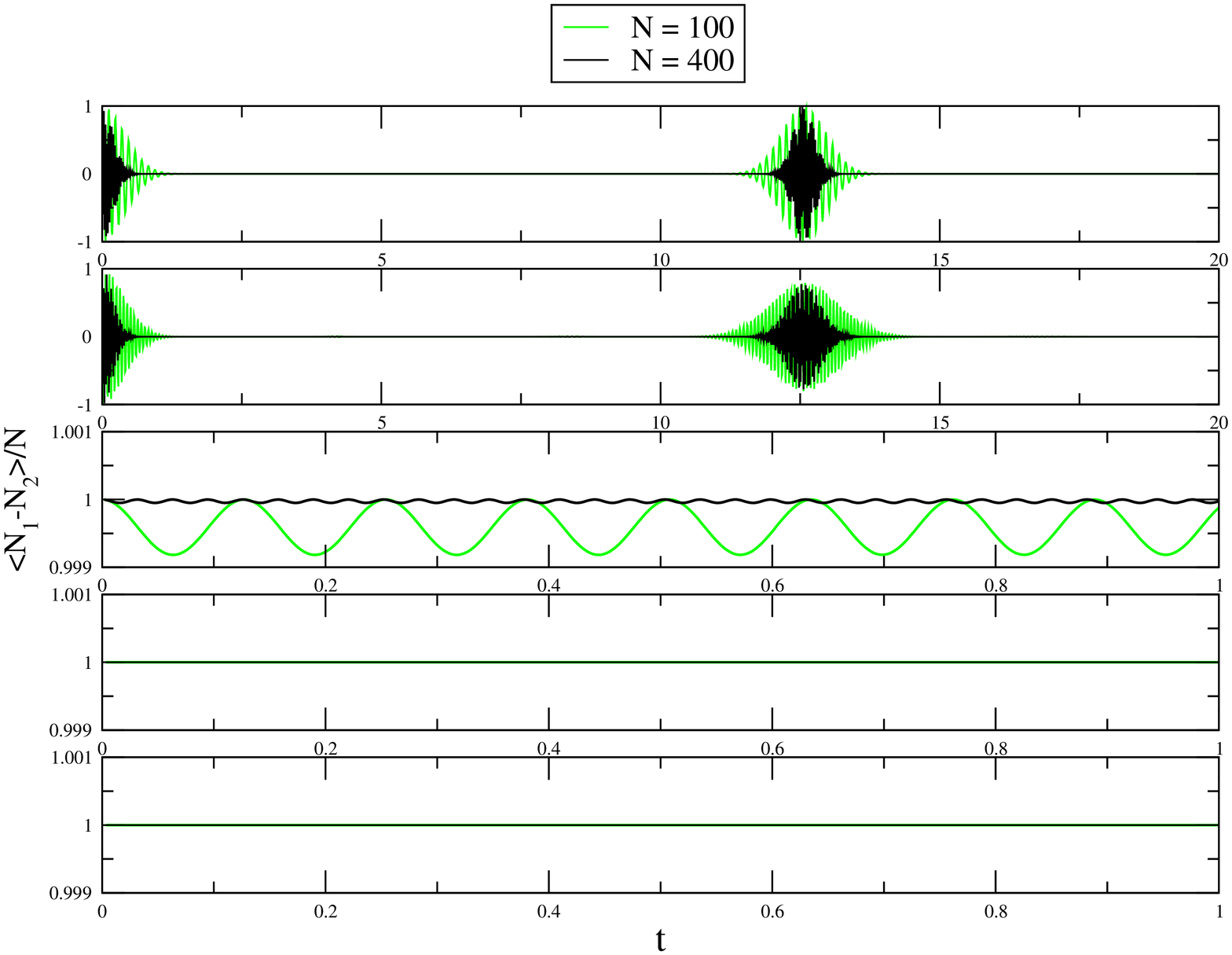}  
\caption{Time evolution of the expectation value for the relative number of particles for different ratios of the coupling
$K/{\cal E}_J$. From top to bottom: $K/{\cal E}_J= 1/N^2,1/N,1,N,N^2$ for $N=100, 400$. 
From \cite{arlei_qd}.}
\label{QD1}
\end{figure}
%%%%%%%%%%%%%%%%%%%%%%%%%%%%%%%%%%%%%%%%%%%%%%%%%%%%%%%%%%%%%%%%%%%%%%%

The model can be exactly solved by the Bethe Ansatz,
allowing analytic computation of physical quantities, such as form factors and correlation
functions \cite{review_Jon} and providing also a 
characterisation of condensate fragmentation in the attractive regime \cite{fragmentation}.
Although very simple, it describes different relevant physical scenarios, such as 
tunneling and  self-trapping,
as can be observed in a study of the quantum dynamics described briefly below.
In figure~\ref{QD1} we show the time evolution of the expectation value of the 
imbalance population $(N_1-N_2)/N$ for different 
ratios of the coupling $K/{\cal E}_J$ and $\Delta \mu=0$. 
The curves in green (black) are for $100~(400)$ particles. 
An initial state is used where all particles are on the left well.

%%%%%%%%%%%%%%%%%%%%%%%%%%%%%%%%%%%%%%%%%%%%%%%%%%%%%%%%%%%%%%%%%%%%%%%
\begin{figure}[ht]
\centering
\includegraphics [width=1.0\linewidth]{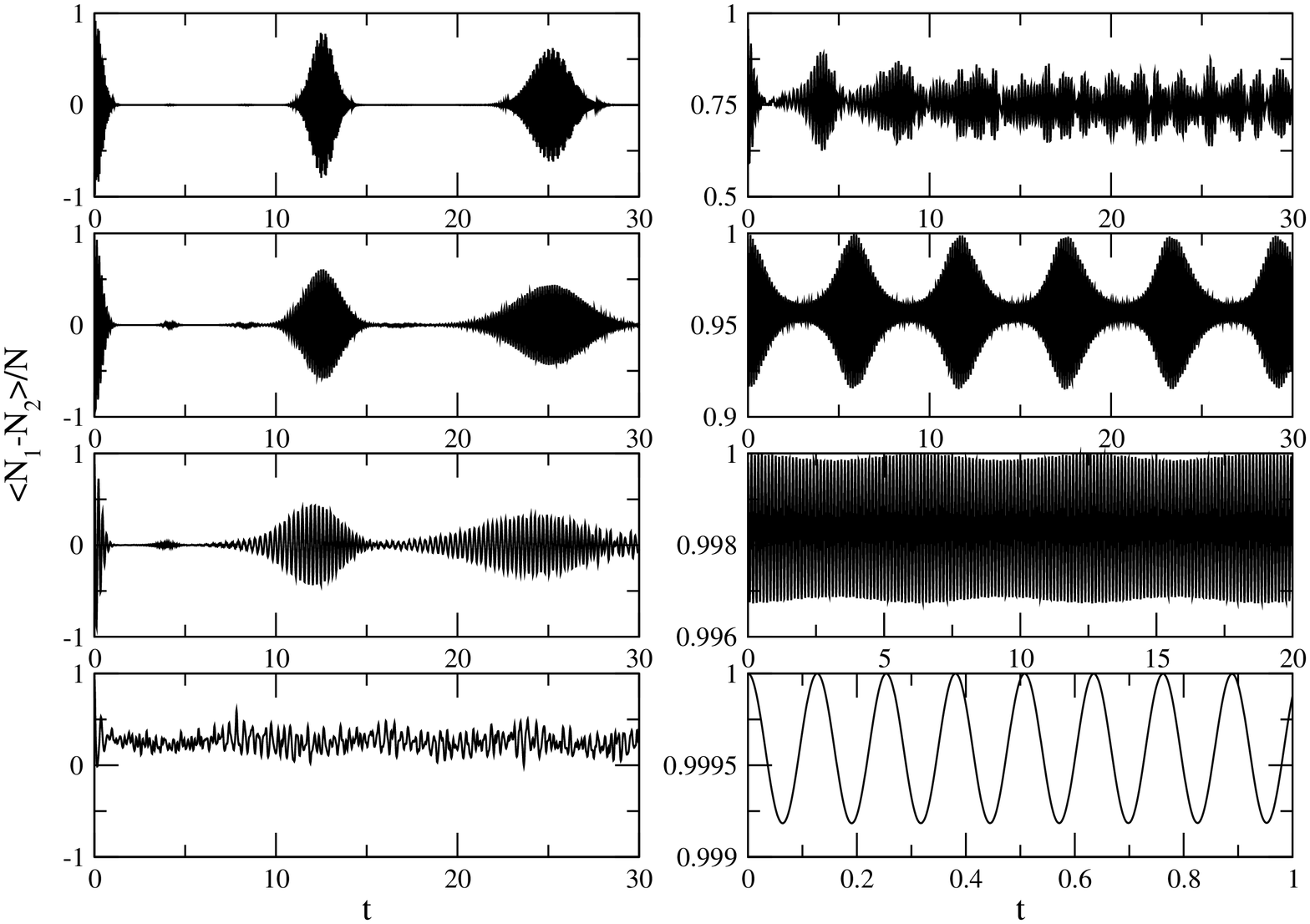}  
\caption{Time evolution of the expectation value between $k/{\cal E}_J=1/N$ and  $k/{\cal E}_J= 1$. 
On the left, from top to bottom $k/{\cal E}_J= 1/N,2/N,3/N,4/N$ and on
the right, from top to bottom $k/{\cal E}_J=5/N,10/N,50/N,1$, where $N=100$. From \cite{arlei_qd}.}
\label{QD2}
\end{figure}
%%%%%%%%%%%%%%%%%%%%%%%%%%%%%%%%%%%%%%%%%%%%%%%%%%%%%%%%%%%%%%%%%%%%%%%

It is clear that the qualitative behaviour does not depend on the precise number of particles.
It is interesting to observe in some plots a collapse and revival behaviour, typical of some experiments.
In the interval between ${K}/{{\cal E}_J}= {1}/{N}$ and ${K}/{{\cal E}_J}=1$ the system tends to localise.
In order to see this better, this region is considered in more detail in figure~\ref{QD2}.
Around ${K}/{{\cal E}_J}={4}/{N}$ there is a threshold in the physical scenario: 
below this value the amplitude of oscillation of the system runs between positive and negative values, 
meaning that the particles can tunnel between the two wells, while for values above this 
threshold the amplitude of oscillation is restricted to positive values, implying that the 
particles are trapped in one of the wells.
Therefore, the quantum dynamics of the two-site Bose Hubbard model describes  
tunnelling and  self-trapping, 
in qualitative agreement with experiments performed in Oberthaler's group \cite{albiez}.

Recently the model has been also used to describe experiments in ultracold 
atomic gases on Einstein-Podolsky-Rosen correlations \cite{oberthaler1} 
and squeezing \cite{oberthaler2}, relevant in the field of quantum metrology. 
It can also be used to describe a quadrupolar nuclei system in Nuclear Magnetic Resonance (NMR) experiments. 
In this context, a discussion about protocols of classical bifurcation has been presented in \cite{RoditiPRA2013}
and a novel implementation of spin-squeezed states in NMR,
carried out in a liquid crystal sample has been conducted in \cite{RoditiPRL2015},
with possible applications in solid state physics.

\section{Concluding remarks and outlook}

We have outlined a number of examples in which Yang-Baxter integrable models are 
relevant to experiments in condensed matter physics and ultracold atoms.
This list should be considered remarkable, not necessarily because of the examples given, 
but arguably also because of what has been omitted. 
For example, we have not touched on the one-dimensional Hubbard model \cite{LW,Hbook,Hexp}.
Nor did we cover the Anderson model and related quantum dot problems \cite{KSL}.
The discrete BCS model is also an important integrable model in condensed matter, 
with solutions due to Richardson and Gaudin of relevance to experiments 
on ultrasmall metallic grains \cite{PhysRep}.
There are a wealth of exactly solved models of this kind which are yet to find their 
way into experiments.\footnote{One only has to glance, for example, through the reprint volume \cite{KE} 
to see just how rich this field is. In addition there 
are interesting families of quantum chains associated with face models which are yet to be fully explored \cite{Pearce}.}
And of course, fundamental work on Yang-Baxter integrable models continues in earnest, with a tremendous 
amount of knowledge accumulated, for example, on correlation functions.
Most recently there has been considerable progress on solving open spin chains with generic integrable 
boundaries \cite{KMN,WBook}.
It is clear that Yang-Baxter integrable models will continue to
offer valuable insights into the description of physical properties and experimental results for decades to come.
In the remainder of this section we present some other relevant themes, providing an outlook for future research.

\subsection{Few-body systems}

Despite their simplicity and importance, the investigation of few-body systems has been challenging \cite{Blume}. 
Further refinement of experimental techniques in cold atoms have allowed the preparation of few-body ensembles of bosons 
\cite{few_bosons} and fermions \cite{jochim_1, jochim_2} in a one-dimensional harmonic trap. 
The experimental realisation of the McGuire impurity model \cite{McGuire_imp} in a one-dimensional Fermi gas 
has also been reported, where the effects of an impurity are measured by increasing the number of 
fermions in the system one by one \cite{Brouzos}.
Other few-particle systems such as the antiferromagnetic Heisenberg spin chain of up to four atoms in a one-dimensional 
trap have also been experimentally realised, allowing the investigation of quantum magnetism in the ultracold few-body  context \cite{jochim_Heisenberg}.\footnote{There have been several proposals to realise spin systems in the setting of cold atoms, 
see, e.g., \cite{Bloch,Lewenstein}. 
Their experimental study thus promises  to span the divide between condensed matter and ultracold atoms.}

These exciting experimental advances stimulated the search for new theoretical methods in few-body systems and also intensified 
the debate on the nature of the crossover from few to many-body physics.
Recent work in this area \cite{zinner} is inspired by questions like {\em When is it
appropriate to consider a system to be few-body}? and {\em When should we think of a system as a true many-body system}? 
In some cases even if a few-body system is not exactly solved by the Bethe Ansatz, some inspiration can still be obtained 
from these methods.
This has been explored, for instance, in \cite{few_1, few_2} where although there is no analytical solution in general, 
the Bethe Ansatz method has been combined with the variational principle to obtain physical quantities,
in agreement with existing results. 
With the prospect of new few-body experiments in mind, 
alternative approaches based on the use of the Bethe Ansatz solution are welcome.

\subsection{Systems out of equilibrium}

The study of quantum systems out of equilibrium is one of the most active frontiers of modern physics.
The remarkable experimental realisation of a quantum Newton cradle \cite{cradle} showing that a one-dimensional
Bose gas of $^{87}$Rb does not thermalise after thousands of collisions generated many discussions in the subject of
integrability versus quantum thermalisation.
Subsequently there has been intense activity in the study of the quantum dynamics of integrable systems,
in particular the behaviour of the prototypical Lieb-Liniger model after a quench \cite{Demler,Caux1}
and a generalised Gibbs ensemble (GGE) has been proposed \cite{Rigol} and improved.
A further contribution in this context was given in \cite{MC12,CE13}  
with the development of the generalised Thermodynamic Bethe Ansatz (GTBA)
approach to nonequilibrium evolution, allowing for the computation 
of local observables at late times after the quench, with impressive 
reduction in computational complexity as compared to previous approaches.
A fundamental issue is the identification of a complete set
of the conserved charges, allowing the GGE to predict the correct steady state properties.
The GGE has been also discussed for other integrable models, 
such as the spin-$\frac12$ Heisenberg XXZ chain \cite{Caux2}, where it was 
shown that the quasi-local conserved charges \cite{Prosen1} are crucial for understanding the 
non-equilibrium dynamics of the system, which generalizes to other integrable models.
Also relevant here is the discovery of Yang-Baxter integrability of boundary driven quantum master equations, 
which has been reviewed recently \cite{Prosen2}.

The observation of the excitation spectrum of the one-dimensional Bose gas \cite{probe} also opens up some 
new possibilities.
It has been suggested that instead of colliding two highly energetic clouds of atoms like in the quantum cradle setting \cite{cradle}, 
an alternative could be to observe the time evolution and potential equilibration of collective excitations via the 
relatively low-energy excitations propagating through the system \cite{probe}.
On a more general note, besides the fundamental and conceptual issues, 
the open problems in the study of quantum systems out of equilibrium
are also of interest to future quantum technologies.

\subsection{Quantum information processing}

Another frontier for experimental developments involving Yang-Baxter integrability is the field of
quantum information and computation, an interdisciplinary area that employs 
the fundamental principles of quantum mechanics to information and computer science \cite{Nielsen,Zheng}.
It has been established \cite{Kauf1, Kauf2} that certain solutions of the Yang-Baxter equation together with local
unitary operators form a universal set of quantum gates, 
following general earlier results \cite{Bry}.
These results have been explored to investigate, for instance, the dynamical evolution of quantum states.
More recently, the Yang-Baxter equation has been connected to teleportation-based quantum
computation \cite{Teleportation}, which is an approach to fault tolerant
quantum computation in which the universal quantum gate set is protected from noise using
the teleportation protocol \cite{Gottesman}.
The applications and possibilities in this area are thus also high.

\subsection{Quantum simulation of the Yang-Baxter equation}

The relevance of Yang-Baxter integrable models in different fields is clear.
Beyond realising further Yang-Baxter integrable systems in the laboratory, 
it is highly desirable to experimentally implement the Yang-Baxter equation itself 
in some setting.
In the optics framework several proposals for simulating the Yang-Baxter equation have been 
discussed \cite{Ge2008}
and a first attempt to verify it has been conducted in \cite{Long2013}
using linear quantum optical components such as beamsplitters,
half-wave plates, quarter-wave plates, etc.
An experimental realisation of the Yang-Baxter equation 
through a Nuclear Magnetic Resonance interferometric setup has been performed on 
a liquid state Iodotrifluoroethylene sample of  molecules containing three 
qubits \cite{Fatima2015},
establishing an additional striking connection between 
integrability on the one hand and quantum information processing on the other.
This opens up a way to implement quantum entanglement with integrability 
and certainly deserves to be further investigated.

\subsection{Artificial spin ice}

Apart from the hard hexagon model, with which we began this article, all of the examples given 
have been for essentially one-dimensional quantum systems. 
Of course, these models have their two-dimensional classical counterparts.
So far there have been, for example, no experiments performed on related two-dimensional vertex models.
As a speculative step in this direction, it may be possible to build on recent experimental advances with 
artificial spin ice.
Artificial spin ice systems are two-dimensional systems constructed with nanomagnetic arrays, 
vortices in nanostructured superconductors, 
or soft matter systems (see, e.g., \cite{spin_ice} and references therein). 
Square ice has also been recently observed with water locked between two graphene sheets \cite{graphene}.
Artificial spin ice can be simulated on the square and kagome lattices via colloids in double-well traps \cite{spin_ice}.
In particular, the systems are built such that the familiar ice-rule is obeyed at each vertex of the square lattice.
Examination of such artificial spin ice within the context of the Yang-Baxter integrable vertex models may not 
be entirely outside the realms of possibility.

\ack
It is a pleasure to thank our numerous colleagues for inspiration and collaboration over the years which has shaped 
our understanding and appreciation of Yang-Baxter integrable models.  
In particular, we mention here Rodney Baxter, Michael Karowski and Xiwen Guan.
We thank Fabrizio Dolcini, Vladimir Gritsev, Jon Links and Tomaz Prosen for helpful comments on this article.
MTB gratefully acknowledges support from Chongqing University and the 1000 Talents Program of China. 
AF acknowledges CNPq (Conselho Nacional de Desenvolvimento Cientifico e Tecnologico) for financial support.
This work has also been partially supported by the Australian Research Council through Discovery Projects 
DP150101294 and DP130102839.

\end{document}